\documentclass[preprint]{sig-alternate-05-2015}

\toappear{}
\usepackage{balance}
\usepackage[table,x11names]{xcolor}
\usepackage{multirow}

\begin{document}






%

\title{Assessing and Comparing Mutation-based Fault Localization Techniques}

%
%
%
%
%

\numberofauthors{3} 
%
\author{
%
%
\alignauthor
Thierry Titcheu Chekam\\
       \affaddr{Interdisciplinary Centre for Security, Reliability and Trust}\\
       \affaddr{University of Luxembourg}\\
       \affaddr{thierry.titcheu-chekam@uni.lu}
\alignauthor
Mike Papadakis\\
       \affaddr{Interdisciplinary Centre for Security, Reliability and Trust}\\
       \affaddr{University of Luxembourg}\\
       \affaddr{michail.papadakis@uni.lu}
\and
\alignauthor 
Yves Le Traon\\
       \affaddr{Interdisciplinary Centre for Security, Reliability and Trust}\\
       \affaddr{University of Luxembourg}\\
       \affaddr{yves.letraon@uni.lu}
}

\maketitle
\begin{abstract}
 Recent research demonstrated that mutation-based fault localization techniques are relatively accurate and practical. However, these methods have never been compared and have only been assessed with simple hand-seeded faults. Therefore, their actual practicality is questionable when it comes to real-wold faults. To deal with this limitation we asses and compare the two main mutation-based fault localization methods, named Metallaxis and MUSE, on a set of real-world programs and faults. Our results based on three typical evaluation metrics indicate that mutation-based fault localization methods are relatively accurate and provide relevant information to developers.  Overall, our result indicate that Metallaxis and MUSE require 18\% and 37\% of the program statements to find the sought faults. Additionally,  both methods locate 50\% and 80\% of the studied faults when developers inspect 10 and 25 statements. 
 
\end{abstract}



\section{Introduction}

Software debugging is an essential software development activity. A typical debugging scenario consists of the following three phases: fault localization, fault comprehension and fault repair. Fault localization refers to the process when developers try to identify statements that they suspect that are problematic. Fault comprehension refers to the process when developers try to understand the actual problem with their code. Once they understand it they try to fix it.  Each one of these phases is painful, costly and thus, research is dedicated to (semi)automate them. 

Fault locazation is the basic step for the above scenario since both fault comprehension and repair are strongly depended on it. Thus, it is natural to expect that more accurate fault localizaton leads to faster fault comprehension and repair. In view of this, spectrum-based fault localization methods have been developed. These methods approximate the likelihood that  program statements are faulty given the execution profile of failed and passing test cases. Then, they rank all the program statements in a decreasing likelihood order in an attempt to guide developers. 

Spectrum-based fault localization techniques have been studied with the use of different spectra types, i.e., statements, branches data-flows, and different likelihood estimation formulas. However despite the efforts of the community, they are relatively inaccurate \cite{ParninO11}. This is partly due to the fact that coverage is not correlated with test effectiveness, which results in the so-called coincidental correctness \cite{MasriA14}, and partly due to the small number of test cases that are typically available. Therefore, it is quite hard, if not impossible, to relate spectra elements with actual faults. Another problem with spectrum-based fault localization is that the ordered statement list contains many statements that are unrelated to the sought fault \cite{ParninO11}.  

The present paper studies the mutation-based fault localization techniques \cite{Papadakis:Metallaxis-FL}, which aim at addressing the above-mentioned limitations of the spectrum-based techniques. The underlying idea of these techniques is to gain information from mutants, which have been shown to correlate with faults \cite{AndrewsBLN06}, and provide guidance regarding the identified faults. Despite the promising results of these methods, today there is no evidence on how well these methods perform in real-world cases, i.e., in localizing real-world faults. This form the primary objective of the this paper. 

In literature, two mutation-based fault localization techniques, named as Metallaxis \cite{Papadakis:Metallaxis-FL} and MUSE \cite{Moon:MUSE},  have been proposed. Metallaxis realises the idea that mutants are coupled with faults (killing mutants results in finding faults) and localizes them based on the mutants that couple with the fault according to the test suite. MUSE is inspired by the automated fault repair tools and localizes faults based on mutants that turn failing executions into passing ones. Given the differences of the two approaches, a natural question to ask is which one is more effective and why. Since, no comparison between them has been attempted a secondary objective of this paper is to compare these techniques and provide insights on when and why one approach performs better than the other.  

For evaluation we use CoREBench, a benchmark of complex real-world faults and show that they provide relatively accurate results. In particular our result indicate that developers using Metallaxis and MUSE successfully locate the studied faults by investigating only 18\% and 37\% of the executable program statements. The results also indicate that Metallaxis was two times less costly than MUSE. 

Our study also reveal that Metallaxis can locate approximately 50\% or 80\% of the studied faults by investigating only 10 or 25 statements, respectively. This is particularly important finding since it ensures that developers will not experience a drop-off on their interest for the fault localization \cite{ParninO11}.

Another major advantage offered by these techniques is that the most suspicious statements are related to the actual faulty program statements. Our result indicate that the average relevance of the top 10 statements is 28.16\% and 25.88\% for Metallaxis and Muse, meaning that on average 3 out of the first 10 statements can lead to the sought fault. This was a key requirement set by the user study of Parin and Orso \cite{ParninO11}. 

Overall, the contributions of the present paper can be summarised on the following points:

\begin{itemize}
\item We evaluate the accuracy of mutation-based fault localisation methods on real-world programs and faults. 
\item We compare and analyse the performance of the mutation-based fault localisation methods.
\item We perform a qualitative analysis of the studied approaches. 
\end{itemize}

The rest of the paper is organized as follows: Sections \ref{sec:Background} and \ref{sec:Study Design} respectively detail the studied approaches and the design of the conducted study. Sections \ref{sec:Quantitative} and \ref{sec:Qualitative} are then analysing the results of the study, in a quantitative and qualitative manner, respectively. Then, Section \ref{threats} and Section \ref{sec:Related-Work} discusses threats to validity and related work. Finally, Section \ref{sec:Conclusion} concludes the paper.

\section{Background} \label{sec:Background}
\subsection{Mutation Analysis}

Mutation  is a software analysis technique that alters (mutates) the syntax of the program under analysis with the intention to produce programs that have small semantic differences from the original one \cite{OffuttH96}. The produced programs are called mutants and are constructed by making simple syntactic transformations on the original program. The transformation rules that are used to produce the mutants are called mutant operators. Mutation has been demonstrated to be quite powerful, mainly due to its ability to force and expose the different behaviours of the program under analysis. A test analysis scenario involves observing the behaviour of the mutants when exercised by test cases. In case a mutant exhibit different behaviour from the original program it is called killed, while it is called live in the opposite case.  

Mutation analysis differs from the coverage-based techniques because it forces the candidate test cases not only to execute specific program locations but also to exercise the sensitivity of these locations to trigger potential errors to the observable output. This requirement of mutation makes it particularly powerful with respect to testing and analysis. Mutants have been shown to be quite effective in mimicking the behaviour of real faults \cite{AndrewsBLN06} and in performing thorough testing that reveal more faults than other test criteria \cite{BakerH13}. The present paper realises the above attributes of mutation, in the context of fault localization, and ultimately examines the ability of mutants to effectively point out the location of real faults.

\subsection{Mutation-based fault localization}

\begin{figure}[t]
\centering
\includegraphics[scale=0.8]{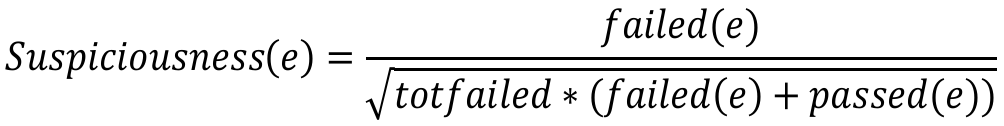}
\vspace{-1.0em}
\caption{The Ochiai formula. In the formula: e represents a mutant, totfailed - the total number of test cases that fail, failed(e) - the number of test cases that kill the mutant e and fail and passed(e) - the number of test cases that kill mutant e and pass.}
\label{fig:Ochiai}
\end{figure}

\begin{figure}[b]
\vspace{-1.0em}
\centering
$\mu(s) = \displaystyle\dfrac{1}{mut(s)}\sum_{m \in mut(s)}\left(\dfrac{|f_{P}(s) \cap p_m|}{|f_{P}|} - \alpha \dfrac{|p_{P}(s) \cap f_m|}{|p_{P}|}\right)$
\vspace{-1.0em}
\caption{The suspiciousness metric of MUSE. In the formula: mut(s) represents the number of mutants residing on the statement s, the term $\frac{|f_{P}(s) \cap p_m|}{|f_{P}|}$ - the proportion of tests that were turned from failing to passing, the term $\frac{|p_{P}(s) \cap f_m|}{|p_{P}|}$ - the proportion tests that were turned from passing to failing, the value a adjusts the average values of the two terms.}
\label{fig:MUSE}
\end{figure}

\begin{figure*}[ht!]
\centering
\includegraphics[scale=1]{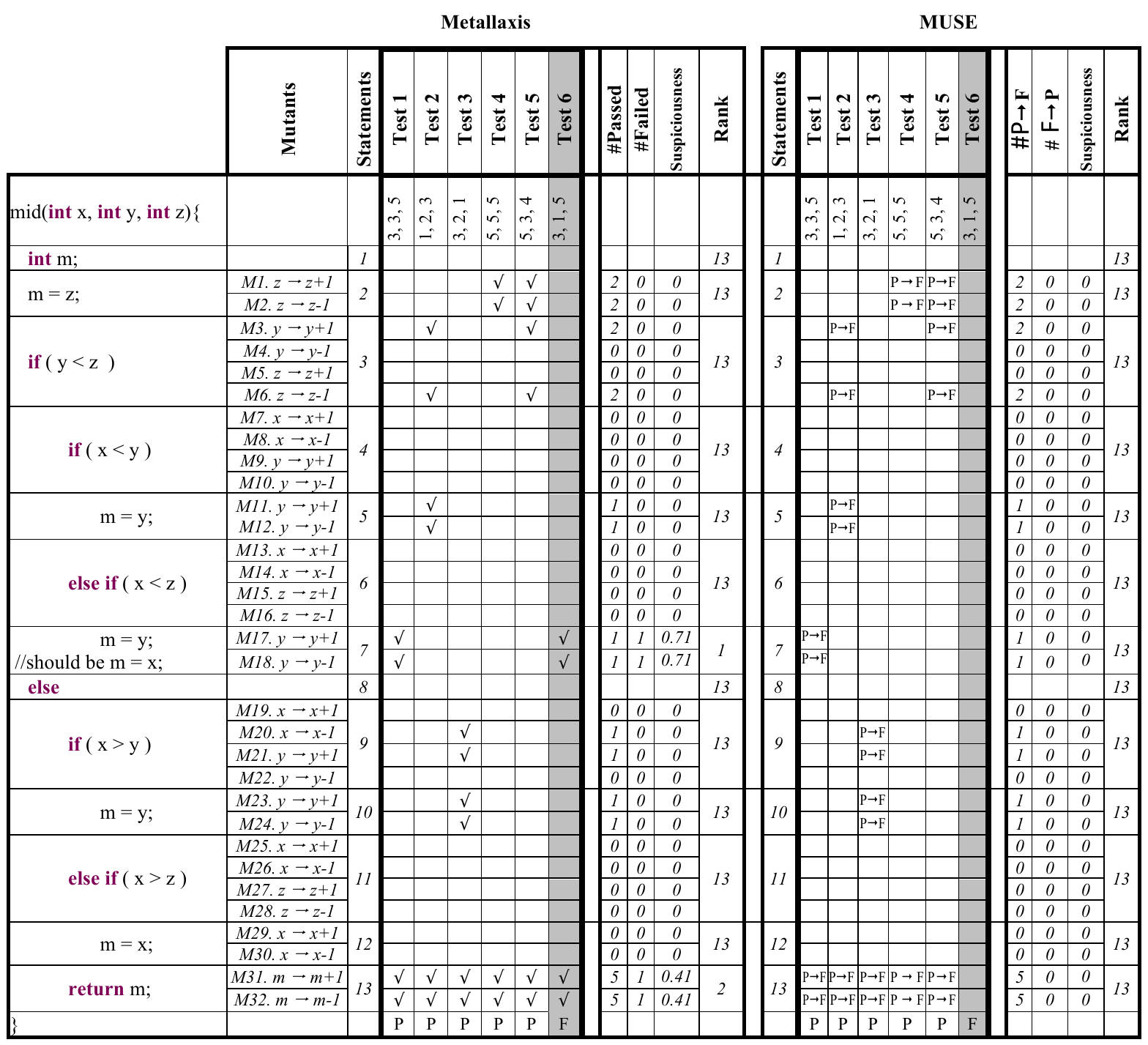}
\vspace{-1.0em}
\caption{Mutation-based fault localisation example when using Metallaxis and MUSE}
\vspace{-1.5em}
\label{fig:FLexample}
\end{figure*}

Fault localization techniques collect dynamic information of the program under analysis and try to associate the executed statements with the experienced failures. In other words they try to identify program statements that are most likely to be faulty. Thus,  for every program statement they estimate and assign a value that represents the probability that it is faulty. This value is called suspiciousness value. To find the faults, users have to inspect the program statements according to their suspiciousness. Thus, fault localization methods produce a priority list of statements that the users have to follow. Statements position in this list is called rank and it is used as a basis of comparison between fault localization methods. 

The underlying idea of fault localization is that entities covered by failing tests are more likely to be responsible for failures than entities covered mostly by passing tests. Mutation-based fault localization realises this idea with mutants. Thus, mutants killed by failing tests and rarely by passing tests are more likely to relate with failures than mutants killed mostly by passing tests. In this paper we are considering two fault localization methods named as Metallaxis \cite{Papadakis:Metallaxis-FL} and MUSE \cite{Moon:MUSE}.

Metallaxis assigns suspiciousness values to the employed mutants using the Ochiai similarity function (shown in figure \ref{fig:Ochiai}). Then, based on the location of the mutants it assigns a suspiciousness value on the program statements (using the maximum value of the included mutants). Additional details regarding Metallaxis can be found in the work of Papadakis and Le Traon \cite{Papadakis:Metallaxis-FL}. 

MUSE follows a different way in identifying suspicious statements. Instead of using the ``traditional" way of judging whether mutants are killed or not (based on program outputs \cite{0020331}), it only considers mutants as killed (or not) based on the passing and failing tests. Thus, MUSE only considers the cases that mutants turn a passing test case to a failing one and vice versa. Thus, it ignores the cases that both mutants and the original (faulty) program fail but in different ways, i.e., their outputs differ.

MUSE assigns suspiciousness values to the program statements using the formula of figure \ref{fig:MUSE}. The first term $\frac{|f_{P}(s) \cap p_m|}{|f_{P}|}$ represents the proportion of tests that were turned from failing to passing. Similarly, the second term $\frac{|p_{P}(s) \cap f_m|}{|p_{P}|}$ of the formula represents the proportion tests that were turned from passing to failing. The value a is used for balancing the two terms and hence it adjusts the average values of the terms. To compute the overall suspiciousness of a statement, the sum of the two terms is divided with the number of mutants residing on the statement. Further details regarding MUSE can be found in the work of Moon et al. \cite{Moon:MUSE}.

\subsection{Example}
Figure \ref{fig:FLexample} presents a working example of how the studied fault localization methods work. The example presents an implementation of a function that takes three integers as input and returns the median value. Assuming that there is a fault on the statement 7 and that the developer discovered it by using 6 test cases, 5 passing and 1 failing. For the shake of the example we will demonstrate the method by using only one mutant operator that adds or subtracts the value of 1 on every program statement that it is relevant. 

Metallaxis works by executing the mutants with the 6 test cases and recording the number of passing and failing tests that kill (causing the program to provide different output) each one of the mutants (marked with `$\surd$'). Then, for every mutant it computes its suspiciousness value using the Ochiai formula \ref{fig:Ochiai}. Finally the statements are assigned with the maximum value of mutant suspiciousness based on which they are ranked. 

MUSE works by measuring the number of tests that are turned from passing to failing and from failing to passing (marked as `P->F' and `F->P'). Then, for every statement a suspiciousness value is assigned according to the formula presented in \ref{fig:MUSE}. Finally the statements are ranked according to their suspiciousness. 

In the example of Figure \ref{fig:FLexample} Metallaxis successfully ranked the faulty statement at the top of its list. This happened since the mutants M17 and M18 were only killed by the test case that failed, i.e., Test6. Unfortunately, since mutants M17 and M18 did not turned the execution of Test6 to passing, MUSE failed to locate the fault. It is noted that Metallaxis considers the contribution of the mutants by observing the output differences between the original and the mutant programs while MUSE by observing the differences between the expected output with that of the original and mutant programs.

\section{Study Design} \label{sec:Study Design}

\subsection{Research Questions}  \label{sec:Research Questions}
The research questions investigated in our empirical study are the following:

 \textbf{RQ1:} How Metallaxis and MUSE compare with the ``optimal'' fault localization?

\textbf{RQ2:} What are the accuracy differences between Metallaxis and MUSE?

\textbf{RQ3:} How relevant are the top ranked statements with the sought faults?

The first research question aims at evaluating the relative accuracy of the mutation-based fault localization techniques, i.e.,  Metallaxis and MUSE, in suggesting the locations, as line number in the source code, when compared to an `optimal'' fault localization. In the past, the two techniques have only been evaluated on hand-seeded faults \cite{Papadakis:Metallaxis-FL, Moon:MUSE} with remarkably good accuracy. However, this might not be the case for the real faults that tend to be more complex than the hand-seeded ones.

The second research question aims at directly comparing the accuracy of Metallaxis and MUSE in recommending the suspicious faulty locations of the studied software faults. The focus here is on the accuracy differences between the two techniques.

The third research question investigates whether statements with the highest suspiciousness values are related to the sought faults. In other words, we seek to investigate whether developers can identify what went wrong by inspecting those statements. In practice localizing a fault requires locating the source code lines that are related and can easily lead to the faulty statement. In fact, some faults are easily understandood after reviewing several other lines related to the faulty ones \cite{ParninO11}. 

\subsection{Subjects}
To conduct the experiment we used the bugs and patches from GNU Coreutils application suite\footnote{http://www.gnu.org/software/coreutils/}, GNU Findutils\footnote{http://www.gnu.org/software/findutils/} and GNU Grep\footnote{http://www.gnu.org/software/grep/} included in the CoREBench suite\footnote{http://www.comp.nus.edu.sg/~release/corebench/} \cite{BohmeR14}.

\begin{table}[ht]
\scriptsize
\centering
\vspace{-1.0em}
\caption{Used subjects and faults.}
\label{table:bugsinfo}
\begin{tabular}{|*{8}{c|}} \cline{2-8} 
\multicolumn{1}{c|}{} &  &  & \textbf{\# of} &  \textbf{Source} & \multicolumn{3}{c|}{ \textbf{\#Tests}} \\ \cline{6-8}
\multicolumn{1}{c|}{} & \multirow{-2}{*}{ \textbf{ID}} &  \multirow{-2}{*}{ \textbf{Prog.}} &  \textbf{SLOC} &  \textbf{File}  &  \textbf{Pass} &  \textbf{Fail} &  \textbf{All}\\ \hline  \hline 
 \multirow{21}{*}{\rotatebox[origin=c]{90}{Coreutils}}
 & 1 & rm  & 1 & rm.c & 70 & 2 & 72 \\ \cline{2-8}
 & 2 & od  & 12 & od.c & 451 & 3 & 454 \\ \cline{2-8}
 & 3 & cut  & 3 & cut.c & 108 & 1 & 109 \\ \cline{2-8}
 & 4 & tail  & 6 & tail.c & 89 & 1 & 90 \\ \cline{2-8}
 & 5 & tail  & 4 & tail.c & 84 & 5 & 89 \\ \cline{2-8}
 & 6 & cut  & 1 & cut.c & 105 & 1 & 106 \\ \cline{2-8}
 & 7 & seq  & 2 & seq.c & 60 & 1 & 61 \\ \cline{2-8}
 & 8 & seq  & 9 & seq.c & 52 & 1 & 53 \\ \cline{2-8}
 & 9 & seq  & 2 & seq.c & 45 & 5 & 50 \\ \cline{2-8}
 & 11 & cut  & 4 & cut.c & 81 & 1 & 82 \\ \cline{2-8}
 & 12 & cut  & 1 & cut.c & 77 & 2 & 79 \\ \cline{2-8}
 & 13 & ls  & 3 & ls.c & 155 & 1 & 156 \\ \cline{2-8}
 & 14 & ls  & 1 & ls.c & 154 & 1 & 155 \\ \cline{2-8}
 & 15 & du  & 1 & du.c & 58 & 2 & 60 \\ \cline{2-8}
 & 16 & tail  & 2 & tail.c & 82 & 1 & 83 \\ \cline{2-8}
 & 17 & cut  & 2 & cut.c & 68 & 3 & 71 \\ \cline{2-8}
 & 18 & seq  & 1 & seq.c & 34 & 1 & 35 \\ \cline{2-8}
 & 19 & seq  & 19 & seq.c & 30 & 1 & 31 \\ \cline{2-8}
 & 20 & seq  & 16 & seq.c & 23 & 7 & 30 \\ \cline{2-8}
 & 21 & cut  & 3 & cut.c & 180 & 2 & 182 \\ \cline{2-8}
 & 22 & expr  & 1 & expr.c & 20 & 2 & 22 \\ \hline \hline

 \multirow{6}{*}{\rotatebox[origin=c]{90}{Findutils}} 
  & 27 & find  & 3 & ftsfind.c & 565 & 1 & 566 \\ \cline{2-8}
 & 32 & find  & 2 & ftsfind.c & 324 & 1 & 325 \\ \cline{2-8}
 & 33 & find  & 2 & ftsfind.c & 324 & 1 & 325 \\ \cline{2-8}
 & 35 & find  & 2 & ftsfind.c & 296 & 1 & 297 \\ \cline{2-8}
 & 36 & find  & 24 & find.c & 24 & 1 & 25 \\ \cline{2-8}
 & 37 & find  & 65 & find.c & 21 & 1 & 22 \\ \hline \hline

 \multirow{3}{*}{\rotatebox[origin=c]{90}{Grep}} 
  & 46 & grep  & 4 & main.c & 1361 & 1 & 1362 \\ \cline{2-8}
 & 47 & grep  & 4 & main.c & 1110 & 1 & 1111 \\ \cline{2-8}
 & 48 & grep  & 27 & main.c & 1364 & 1 & 1365 \\ \hline
\end{tabular}
\end{table}

\textbf{Faults and Programs.} CoREBench contains 22, 15 and 15 pairs of \{\textit{bug-introduction, bug-fixing}\} versions of the Coreutils, Findutils and Grep. We consider bugs whose bug-fixes were made in the source file containing the C language \texttt{main} function; such selection result in a total of \textbf{$\mathbf{30}$ bugs}. The executable source code lines (SLOC) changed during bug-fixing commit are considered as bug-fixes and chosen to be \textit{Fault Locations}. For each bug (\textit{B}), the newest version of the corresponding program preceding the bug-fixing commit is chosen as subject program (\textit{P}). 

\textbf{Test Suites.} For each bug \textit{B} with corresponding faulty program \textit{P}, the developer test suite included in \textit{P}'s repository is used in the experiments. We augment such test suite with the regression test cases added by \textit{B}'s bug-fixing commit and the regression tests shipped with CoREBench to test for \textit{B}. We found that most of the test cases included in \textit{P}'s test suite are large script files containing multiple smaller test cases. This is a common phenomenon in practice which hinders the effectiveness of fault localization \cite{XuanM14}. Therefore, we follow the suggestions of Xuan and Monperrus \cite{XuanM14} and manually splitted such tests in order to have finer granularity and therefore more accuracy in the results.

Table \ref{table:bugsinfo} records information about the used subjects. In this table, the first column records the CoREBench ID for the bug. The column \textit{Prog.} the name of the buggy program, the column \textit{\#Faulty SLOC} the number of executable lines of code changed in the bug-fixing commit\footnote{Code insertion is represented by change in both the lines preceding and following the insertion point}. The column \textit{Source File} records the source file containing the definition of the C language \textit{main} function of the corresponding buggy program. The columns \textit{Pass} and \textit{Fail} respectively show,  for a subject,  the number of tests that pass and fail on the buggy program due to the corresponding fault. The column \textit{All} records the number of tests cases for each subject.

\subsection{Mutation Operators}
\label{operators}
To perform the mutation analysis we used the mutation testing tool that was developed and used in the studies of Henard et al. \cite{CPHJT16}. 
 The tool supports the following mutant operators: Arithmetic (AOR), Logical Connector Replacement (LCR), Relational (ROR), Unary Operator Mutation (UOM), Arithmetic assignment mutation (OAAA), Bitwise operator mutation (OBBN), Logical context negation (OCNG), Statement Deletion (SSDL) and Integer Constant replacement (CRCR).

\subsection{Experiment Design}

\subsubsection{Procedure}
For each bug \textit{B} with corresponding faulty program \textit{P}, the expriment procedure follows:
\begin{enumerate}
	\item Construct line coverage matrix. The line coverage matrix has one row for each test case and one column for each line of code in the preprocessed source file. Entry $(i,j)$ shows whether test $i$ covered line $j$. 
	\item Generate mutants and remove duplicates. First order mutation is applied on the statements of the source file presented in Table \ref{table:bugsinfo} in order to generate mutants of the buggy program. Mutation testing tools tend to generate many duplicate mutants. Since these can be numerous, approximately 20\% \cite{Papadakis:Trivial}, they can artificially increasing the number of mutants and potentially influencing the fault localization results. Thus, we eliminated them using the TCE approach\cite{Papadakis:Trivial}. The number of mutants generated and remaining after duplicates removal for each subject is shown in Table \ref{table:mutantinfo} where the column \textit{ID} shows the CoREBench ID for the bug. The column \textit{Gen} shows the number of mutants generated by the mutation tool. The columns \textit{Dup.} and \textit{No-Dup.} respectively show the number of duplicate mutants and the number remaining after duplicates removal (non-duplicates). The columns \textit{Live} shows the number of non-duplicates mutants equivalent to the original program with respect to the tests suite (dormant mutants). Finally, the column \textit{Killed (MS)} shows the mutants whose test result differ from the original program for at least one test case (non-dormant mutants) and the respective mutation score. In the remaining part of this paper, we use the word mutants to refer non-dormant mutants.
	\item Generate a result matrix.  All the tests cases are both executed with the program \textit{P} prior mutation and with each one of the mutants. The resulting matrix has one row for each mutant and one column for each test case. Entry $(i,j)$ shows whether test $i$ passed or failed on mutant $j$.
	\item Execute both Metallaxis and MUSE.  We implement the methods in Python programming language. In our implementation, the techniques use the data from the line coverage matrix, results matrix and the mutants information to generate a ranking of the SLOC according the suspiciousness of being faulty.
	\item Gather line number of faulty SLOC.  Regarding RQs 1-4, locations of each faulty line are obtained by manually reviewing the source code changes (``diff") from the bug-fix commit report accessible through CoREBench webpage. Note that the locations of the changed SLOC are obtained after source code pre-processing with a compiler, thus a single SLOC change in a bug-fixing commit may result in several faulty lines gathered. This may happen when the SLOC's changes appear in a C language macro. Changes where a new function definition is added are not considered, the reason being that such a change is not representative of the fault locations.
\end{enumerate}

\begin{table}
\scriptsize
\centering
\vspace{-1.0em}
\caption{Mutants used.}
\label{table:mutantinfo}
\begin{tabular}{|c|c|c|c|c|c|} \hline 
 &  \multicolumn{5}{c|}{\textbf{\# of Mutants}}  \\ \cline{2-6}
\multirow{-2}{*}{ \textbf{ID}} &\textbf{Gen.} & \textbf{Dup. } & \textbf{No-Dup.} & \textbf{Live} & \textbf{\tiny{Killed (MS)}} \\ \hline \hline
1 & 1531 & 1203 & 328 & 177 & \cellcolor[gray]{0.9} 151 (10\%) \\ \hline 
2 & 5785 & 2496 & 3289 & 1778 & \cellcolor[gray]{0.9} 1511 (26\%) \\ \hline 
3 & 2749 & 1490 & 1259 & 963 & \cellcolor[gray]{0.9} 296 (11\%) \\ \hline 
4 & 6413 & 2687 & 3726 & 1956 & \cellcolor[gray]{0.9} 1770 (28\%) \\ \hline 
5 & 6399 & 2682 & 3717 & 1970 & \cellcolor[gray]{0.9} 1747 (27\%) \\ \hline 
6 & 2692 & 1492 & 1200 & 368 & \cellcolor[gray]{0.9} 832 (31\%) \\ \hline 
7 & 2530 & 1349 & 1181 & 355 & \cellcolor[gray]{0.9} 826 (33\%) \\ \hline 
8 & 2364 & 1200 & 1164 & 379 & \cellcolor[gray]{0.9} 785 (33\%) \\ \hline 
9 & 2322 & 1189 & 1133 & 390 & \cellcolor[gray]{0.9} 743 (32\%) \\ \hline 
11 & 2512 & 1329 & 1183 & 372 & \cellcolor[gray]{0.9} 811 (32\%) \\ \hline 
12 & 2659 & 1580 & 1079 & 388 & \cellcolor[gray]{0.9} 691 (26\%) \\ \hline 
13 & 12262 & 4611 & 7651 & 4808 & \cellcolor[gray]{0.9} 2843 (23\%) \\ \hline 
14 & 12238 & 4569 & 7669 & 4806 & \cellcolor[gray]{0.9} 2863 (23\%) \\ \hline 
15 & 2435 & 1382 & 1053 & 518 & \cellcolor[gray]{0.9} 535 (22\%) \\ \hline 
16 & 5472 & 2232 & 3240 & 709 & \cellcolor[gray]{0.9} 2531 (46\%) \\ \hline 
17 & 2405 & 1262 & 1143 & 370 & \cellcolor[gray]{0.9} 773 (32\%) \\ \hline 
18 & 1620 & 927 & 693 & 209 & \cellcolor[gray]{0.9} 484 (30\%) \\ \hline 
19 & 1409 & 793 & 616 & 219 & \cellcolor[gray]{0.9} 397 (28\%) \\ \hline 
20 & 1369 & 773 & 596 & 210 & \cellcolor[gray]{0.9} 386 (28\%) \\ \hline 
21 & 2108 & 1003 & 1105 & 373 & \cellcolor[gray]{0.9} 732 (35\%) \\ \hline 
22 & 1729 & 556 & 1173 & 602 & \cellcolor[gray]{0.9} 571 (33\%) \\ \hline \hline
27 & 1428 & 594 & 834 & 531 & \cellcolor[gray]{0.9} 303 (21\%) \\ \hline 
32 & 1362 & 567 & 795 & 556 & \cellcolor[gray]{0.9} 239 (18\%) \\ \hline 
33 & 1353 & 563 & 790 & 550 & \cellcolor[gray]{0.9} 240 (18\%) \\ \hline 
35 & 970 & 412 & 558 & 352 & \cellcolor[gray]{0.9} 206 (21\%) \\ \hline 
36 & 2570 & 897 & 1673 & 1058 & \cellcolor[gray]{0.9} 615 (24\%) \\ \hline 
37 & 2665 & 923 & 1742 & 1132 & \cellcolor[gray]{0.9} 610 (23\%) \\ \hline  \hline
46 & 5450 & 1767 & 3683 & 1752 & \cellcolor[gray]{0.9} 1931 (35\%) \\ \hline 
47 & 5240 & 1755 & 3485 & 1894 & \cellcolor[gray]{0.9} 1591 (30\%) \\ \hline 
48 & 5459 & 1770 & 3689 & 1741 & \cellcolor[gray]{0.9} 1948 (36\%) \\ \hline 
\end{tabular}
\end{table}

Several SLOCs may have the same suspiciousness value. Simply sorting the SLOCs by suspiciousness value would give different ranking for SLOC with same suspiciousness value. In this experiment, for such situation, we use an approach that is commonly used in literature \cite{Papadakis:Metallaxis, Jones:Score, Jeffrey:2008:FLU}; were all SLOC having same suspiciousness value are ranked together at the upper of their ranks (\textit{Upper-Rank}). We extend this with regard to the situation where more than one faulty SLOC are among the SLOC with same suspiciousness value $v$ as following: all SLOC with suspiciousness value $v$ are ranked at the rank they would have using \textit{Upper-Rank} when only one faulty SLOC has suspiciousness value $v$. Altogether, suppose we have $n$ SLOC having suspiciousness value $v$ among which $k, 0\leqslant k \leqslant n$ are faulty and $n-k$ non-faulty; the lower rank $i$ and upper rank $j=i+n-1$. All these $n$ SLOC having suspiciousness value $v$ are ranked at rank $r=j-k+1$. The reason of this choice follow from the assumption that locating any faulty SLOC is equivalent to localize the fault, assumption made by some metrics used in this study.

\subsubsection{Metrics} \label{sec:Metrics}
To assess the effectiveness of the studied fault localization techniques we adopt the following three metrics, which are common in literature, e.g., \cite{Le:MAP-TopN}:
\begin{itemize}
\item \textbf{Top N: } Measures the number of faults found by an FL technique when considering the \textit{top-N} results in its ranking \cite{Saha:TopN, Zhou:TopN, Le:MAP-TopN}. For a fault affecting several lines of code, if one of the faulty lines is found in the \textit{top-N} results is considered localized.
\item \textbf{Percentage Score (PS): } Measures the percentage of SLOC that are not checked by a programmer reviewing the SLOC in decreasing suspiciousness order until she reaches any faulty line. this ``\textit{score}" measure was proposed in the empirical evaluation of the tarantula automatic FL system by Jones et al. \cite{Jones:Score}. PS is calculated as:
\begin{displaymath}
PS=\frac{\mbox{\# of ranked SLOCs}-rank}{ \mbox{\# of ranked SLOCs}},
\end{displaymath}
were $rank$ denote the the minimum of the ranks of all faulty SLOCs.
 The mean of PSs (MPS) across all bugs among the subjects is also used as metric in this study. 
\item \textbf{Mean Average precision (MAP): } Measures the mean of average precision for an FL technique with all the bugs considered in the study \cite{Le:MAP-TopN, Manning:MAP}. An average precision is computed for each bug, as following \cite{Le:MAP-TopN}:
\begin{displaymath}
AP=\sum_{k=1}^{M}\frac{P(k)\times pos(k)}{\mbox{number of faulty lines}}
\end{displaymath} 
In the above formula, $k$ is a rank in the returned ranked SLOCs, $M$ is the number of ranked SLOCs, $pos(k)$ is a flag showing whether the $k\textsuperscript{th}$ SLOC is faulty or not, respectively taking values $1$ and $0$. $P(k)$ is the precision considering a given \textit{top-k} SLOC, and computed with the formula:
\begin{displaymath}
P(k)=\frac{\mbox{\# faulty SLOC in the top-k}}{k}
\end{displaymath}

Note that when several SLOCs have the same suspiciousness value, we may have $P(k)>1$.
\end{itemize}

In this experiment and regarding RQ1 we compare with the hypothetical ``optimal'' fault technique which is: for each fault, \textit{all faulty lines are ranked with the highest suspiciousness}.

\subsubsection{Utilized Tools}
In the present study, we employed the mutation testing tool developed and used in the studies of Henard et al. \cite{CPHJT16}. 
 The tool is built on top of the program matching and transformation framework called Coccinelle \cite{PadioleauLHM08}. It operates at the source code level by matching pattern instances described in a dedicated language called semantic patch language. GNU Gcov\footnote{Gcov is a test coverage program part of GNU GCC} is used to generate the line coverage matrix. We encoutered some issues while using Gcov on our subjects: several bugs lead \textit{Segmentation Fault} or \textit{Infinite Loop} when executing regression tests, which cause the regression test execution to be ``killed". Therefore, Gcov is not able to flush the coverage data to the disk. This scenario could create loss of information about the fault location in MUSE. In order to tackle such problem, we executed the regression test case with the project debugger GNU GDB\footnote{https://www.gnu.org/software/gdb/}. Once the faulty program is killed, we manually call Gcov's \textit{\_\_gcov\_flush()} function in GDB to flush the coverage data obtained before \textit{P} was ``killed".

\section{Quantitative Analysis}
\label{sec:Quantitative}

\subsection{Practicality of Mutation-based FL}

To answer RQ1 we assess the accuracy of Metallaxis and MUSE. We use the metrics presented in Section \ref{sec:Metrics} to evaluate them with regard to the hypothetical ``\textit{optimal}" technique. Tables \ref{table:accuracy} and \ref{table:AP-PS by Bug} record the performance results of Metallaxis, MUSE and that of the \textit{optimal} technique in terms of the used metrics, i.e., Top N, MPS, MAP, AP and PS. Statistics regarding the faults that are hard and easy to localize are presented in Table \ref{table:Easy and Hard to localize}. 

Table \ref{table:accuracy} presents a summary of the Top N metric value for each one of the FL technique. Each cell represents the value of the corresponding metric (row) for the corresponding technique (column). The results show that among all the faults, around $16\%$ of them are ranked at the first position and $50\%$ among the top $10$ by both Metallaxis and MUSE. This shows the practicallity of both techniques to localize real faults, despite the high complexity of the studied faults, as demonstrated in the CoREBench authors \cite{BohmeR14}. Similarly, the MPS values are high and indicate the a developer following their faulty SLOC localization ranking would not need to check 82\% of the code when using Metallaxis and 63\% of the code when using MUSE.    
Both Metallaxis and MUSE have a very low MAP value, which again indicates that both techniques are relatively accurate. 

\begin{table}
\small
\centering
\caption{Accuracy Assesment.}
\label{table:accuracy}
\begin{tabular}{|c||c|>{\columncolor[gray]{0.9}}c|c|} \hline
\textbf{Metrics} & \textbf{Metallaxis} & \textbf{\textit{Optimal}} & \textbf{MUSE} \\ \hline \hline
\textbf{Top 1} & 3 & 30 & 4 \\ \hline 
\textbf{Top 5} & 9 & 30 & 8 \\ \hline 
\textbf{Top 10} & 15 & 30 & 14 \\ \hline 
\textbf{Top 15} & 17 & 30 & 15 \\ \hline 
\textbf{Top 20} & 19 & 30 & 16 \\ \hline 
\textbf{Top 25} & 21 & 30 & 16 \\ \hline 
\textbf{Top 30} & 24 & 30 & 18 \\ \hline 
\textbf{Top 35} & 24 & 30 & 19 \\ \hline 
\textbf{MPS} & 0.819 & 0.994 & 0.628 \\ \hline 
\textbf{MAP} & 0.162 & 1.000 & 0.160 \\ \hline
\end{tabular}
\end{table}

\textbf{Faults Easy and Hard to Localize.} Table \ref{table:Easy and Hard to localize} shows for both Metallaxis (left side) and MUSE (right side), the subjects whose faults are easily localized (top end of the table) and those hard to localize (bottom end of the table). For each technique, the rank of the top ranked faulty SLOCs is shown. We consider a fault \textit{F} easy to localize by a technique \textit{T} when \textit{F} is ranked by \textit{T} among the top $10$ and have high PS and AP values (relative to the subject analysed). Similarily, \textit{F} is hard to localize by \textit{T} when \textit{F} rank is greater than $10$ and very low PS and AP values. Table \ref{table:AP-PS by Bug} shows AP and PS values for each Fault. From both Tables \ref{table:AP-PS by Bug} and \ref{table:Easy and Hard to localize}, we observe that (1) Faults having both high AP ($\geqslant 0.1$) and high PS ($\geqslant 0.7$) are ranked to the high (top 10). And (2) fault having low AP ($\leqslant 0.99$) and high PS ($\geqslant 0.7$) are ranked after the top 10 but within the top 50. (3) The remaing have relatively low AP and PS. 

We consider for this study the fauls of type (1) from the previous paragraph, as \textit{easy to localize}, those of type (3) \textit{hard to lolocalize}, and those of type (3) as \textit{average}. Roughly $50\%$ of the $30$ faults are easily found, respectively $13\%$ and  $37\%$ hard to localize by Metallaxis and MUSE. 

\begin{table}
\small
\centering
\caption{AP and PS values for the considered faults.}
\label{table:AP-PS by Bug}
\begin{tabular}{|c||c|c|c c|c c|} \hline
& &  & \multicolumn{2}{c|}{\textbf{AP}} & \multicolumn{2}{c|}{\textbf{PS}} \\ \cline{4-7}
\multirow{-2}{*}{}&\multirow{-2}{*}{\textbf{ID}}&\multirow{-2}{*}{\textbf{Prog.}}&\textbf{Meta.}&\textbf{MUSE}&\textbf{Meta.}&\textbf{MUSE} \\ \hline \hline
$l_{1}$&1&rm&\textbf{1.000}&\textbf{1.000}&\textbf{0.984}&\textbf{0.984}\\ \hline 
$l_{2}$&15&du&\textbf{1.000}&0.167&\textbf{0.995}&0.969\\ \hline 
$l_{3}$&35&find&\textbf{1.000}&\textbf{1.000}&\textbf{0.984}&\textbf{0.984}\\ \hline 
$l_{4}$&11&cut&0.312&\textbf{0.612}&0.991&\textbf{0.995}\\ \hline 
$l_{5}$&6&cut&\textbf{0.200}&0.005&\textbf{0.977}&0.033\\ \hline 
$l_{6}$&46&grep&\textbf{0.198}&0.119&\textbf{0.987}&0.981\\ \hline 
$l_{7}$&22&expr&0.167&\textbf{0.500}&0.959&\textbf{0.986}\\ \hline 
$l_{8}$&5&tail&\textbf{0.159}&0.037&\textbf{0.993}&0.987\\ \hline 
$l_{9}$&7&seq&\textbf{0.156}&0.127&\textbf{0.971}&0.949\\ \hline 
$l_{10}$&47&grep&\textbf{0.125}&\textbf{0.125}&\textbf{0.991}&\textbf{0.991}\\ \hline 
$l_{11}$&27&find&0.075&\textbf{0.151}&0.867&\textbf{0.933}\\ \hline 
$l_{12}$&14&ls&0.053&\textbf{0.091}&0.979&\textbf{0.988}\\ \hline 
$l_{13}$&12&cut&\textbf{0.050}&0.030&\textbf{0.970}&0.955\\ \hline 
$l_{14}$&9&seq&\textbf{0.050}&0.003&\textbf{0.939}&0.037\\ \hline 
$l_{15}$&33&find&\textbf{0.050}&0.017&\textbf{0.865}&0.595\\ \hline 
$l_{16}$&8&seq&\textbf{0.049}&0.004&\textbf{0.871}&0.064\\ \hline 
$l_{17}$&19&seq&0.041&\textbf{0.111}&0.969&\textbf{0.990}\\ \hline 
$l_{18}$&20&seq&0.037&\textbf{0.307}&0.870&\textbf{0.967}\\ \hline 
$l_{19}$&18&seq&0.037&\textbf{0.333}&0.763&\textbf{0.974}\\ \hline 
$l_{20}$&21&cut&\textbf{0.035}&0.005&\textbf{0.899}&0.040\\ \hline 
$l_{21}$&36&find&\textbf{0.016}&0.008&\textbf{0.952}&0.899\\ \hline 
$l_{22}$&3&cut&0.014&\textbf{0.020}&0.489&\textbf{0.707}\\ \hline 
$l_{23}$&17&cut&\textbf{0.014}&0.003&\textbf{0.818}&0.039\\ \hline 
$l_{24}$&48&grep&\textbf{0.012}&0.001&\textbf{0.952}&0.085\\ \hline 
$l_{25}$&4&tail&\textbf{0.008}&0.002&\textbf{0.948}&0.199\\ \hline 
$l_{26}$&32&find&0.007&\textbf{0.007}&0.000&\textbf{0.041}\\ \hline 
$l_{27}$&2&od&\textbf{0.004}&0.003&\textbf{0.705}&0.363\\ \hline 
$l_{28}$&16&tail&0.001&\textbf{0.001}&0.000&\textbf{0.214}\\ \hline 
$l_{29}$&37&find&\textbf{0.001}&0.001&\textbf{0.874}&0.806\\ \hline 
$l_{30}$&13&ls&0.000&\textbf{0.000}&0.000&\textbf{0.080}\\ \hline
\end{tabular}
\end{table}

\begin{table}
\small
\centering
\caption{Faults Easy and Hard to localize.}
\label{table:Easy and Hard to localize}
\begin{tabular}{|l||c|c|c|||c|c|c|} \hline
&\textbf{ID} & \textbf{Prog.} & \textbf{Meta.} & \textbf{ID} & \textbf{Prog.}  & \textbf{MUSE} \\ \hline \hline
$l_1$&1 & rm & 1 & 1 & rm & 1 \\ \hline
$l_2$&15 & du & 1 & 11 & cut & 1 \\ \hline
$l_3$&35 & find & 1 & 19 & seq & 1 \\ \hline
$l_4$&11 & cut & 2 & 35 & find & 1 \\ \hline
$l_5$&19 & seq & 3 & 22 & expr & 2 \\ \hline
$l_6$&5 & tail & 4 & 18 & seq & 3 \\ \hline
$l_7$&47 & grep & 4 & 20 & seq & 3 \\ \hline
$l_8$&6 & cut & 5 & 47 & grep & 4 \\ \hline
$l_9$&7 & seq & 5 & 15 & du & 6 \\ \hline
$l_{10}$&12 & cut & 6 & 27 & find & 6 \\ \hline
$l_{11}$&22 & expr & 6 & 5 & tail & 7 \\ \hline
$l_{12}$&46 & grep & 7 & 7 & seq & 9 \\ \hline
$l_{13}$&36 & find & 9 & 12 & cut & 9 \\ \hline
$l_{14}$&9 & seq & 10 & 46 & grep & 10 \\ \hline
$l_{15}$&33 & find & 10 & 14 & ls & 11 \\ \hline
$l_{16}$&20 & seq & 12 & 36 & find & 19 \\ \hline
$l_{17}$&27 & find & 12 & 3 & cut & 27 \\ \hline
$l_{18}$&14 & ls & 19 & 33 & find & 30 \\ \hline
$l_{19}$&21 & cut & 20 & 37 & find & 34 \\ \hline
$l_{20}$&8 & seq & 22 & 32 & find & 70 \\ \hline
$l_{21}$&37 & find & 22 & 9 & seq & 157 \\ \hline
$l_{22}$&48 & grep & 26 & 8 & seq & 160 \\ \hline
$l_{23}$&18 & seq & 27 & 2 & od & 186 \\ \hline
$l_{24}$&4 & tail & 28 & 21 & cut & 191 \\ \hline
$l_{25}$&17 & cut & 37 & 17 & cut & 195 \\ \hline
$l_{26}$&3 & cut & 47 & 6 & cut & 206 \\ \hline
$l_{27}$&32 & find & 73 & 4 & tail & 431 \\ \hline
$l_{28}$&2 & od & 86 & 16 & tail & 447 \\ \hline
$l_{29}$&16 & tail & 569 & 48 & grep & 497 \\ \hline
$l_{30}$&13 & ls & 905 & 13 & ls & 833 \\ \hline
\end{tabular}
\end{table}

\subsection{Metallaxis vs MuSE}

In this section, we present the results of our comparison in an attempt to answer RQ2.

\subsubsection{Comparison using Different Metrics}
\begin{figure}[t]
\includegraphics{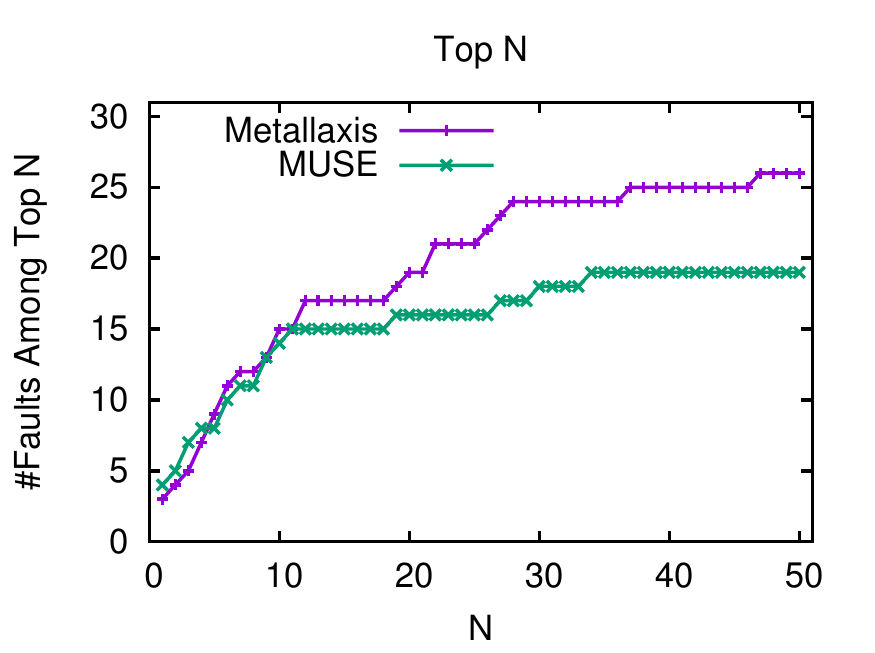}
\caption{Top N, Metallaxis vs MUSE }
\label{fig:Meta-Muse-TopN}
\end{figure}

Figure \ref{fig:Meta-Muse-TopN} visualises the results of the comparison regarding the Top N metric. From these data we can observe that when $N \leqslant 5$, MUSE localize 1 fault more than Metallaxis, when $4\geqslant N \geqslant 11 $ the two technique perform the same, although the $top-k$ localized faults are not same for both techniques as shown in Table \ref{table:Easy and Hard to localize}. As $N$ grows, the number of faults localized by Metallaxis increase faster than MUSE's, showing that Metallaxis is more stable than MUSE. Furthermore, referring to Table \ref{table:accuracy}, Metallaxis have both higher MAP and MPS values than MUSE. 

\subsubsection{Comparison Per-Fault Ranking Difference}

Table \ref{table:Ranking Difference per Fault} records a side by side rank that is given by each one of the considered faults and techniques. The table also records the difference between the rank given by Metallaxis and that of MUSE (rank difference). Over the $30$ faults, $11$ faults (\textit{common}) are ranked in the top $10$ both by Metallaxis and MUSE (middle section of Table \ref{table:Ranking Difference per Fault}: $l_8$ and $l_{10}\sim l_{19}$),  $3$ faults (\textit{MUSE-Metallaxis}) ranked in the top $10$ by MUSE and not Metallaxis (upper section of Table \ref{table:Ranking Difference per Fault}: $l_3$, $l_5$ and $l_{7}$), and finally $4$ faults (\textit{Metallaxis-MUSE}) are ranked in top $10$ by Metallaxis and not MUSE (lower section of Table \ref{table:Ranking Difference per Fault}: $l_{20}$, $l_{22}$, $l_{25}$ and $l_{28}$). The absolute difference in the rankings is not greater than $5$ for all \textit{common} faults, but greater than $5$ for both \textit{Metallaxis-MUSE} and \textit{MUSE-Metallaxis}. In the next section, we investigate \textit{common}, \textit{MUSE-Metallaxis} and \textit{Metallaxis-MUSE} faults.

\begin{table}
\small
\centering
\caption{Per-Fault Ranking Difference.}
\label{table:Ranking Difference per Fault}
\begin{tabular}{|l||c|c|c|c|c|} \hline
 && & \multicolumn{3}{c|}{\textbf{Highest Fault Rank}} \\ \cline{4-6}
\multirow{-2}{*}{}&\multirow{-2}{*}{\textbf{ID}}&\multirow{-2}{*}{\textbf{Prog.}}&\textbf{Metallaxis}&\textbf{MUSE}&\textbf{Diff}\\  \hline \hline
$l_1$&16&tail&569&447&\cellcolor[gray]{0.9} 122\\ \hline 
$l_2$&13&ls&905&833&\cellcolor[gray]{0.9} 72\\ \hline 
$l_3$&18&seq&27&3&\cellcolor[gray]{0.9} 24\\ \hline 
$l_4$&3&cut&47&27&\cellcolor[gray]{0.9} 20\\ \hline 
$l_5$&20&seq&12&3&\cellcolor[gray]{0.9} 9\\ \hline 
$l_6$&14&ls&19&11&\cellcolor[gray]{0.9} 8\\ \hline 
$l_7$&27&find&12&6&\cellcolor[gray]{0.9} 6\\ \hline \hline
$l_8$&22&expr&6&2&\cellcolor[gray]{0.9} 4\\ \hline 
$l_9$&32&find&73&70&\cellcolor[gray]{0.9} 3\\ \hline 
$l_{10}$&19&seq&3&1&\cellcolor[gray]{0.9} 2\\ \hline 
$l_{11}$&11&cut&2&1&\cellcolor[gray]{0.9} 1\\ \hline 
$l_{12}$&1&rm&1&1&\cellcolor[gray]{0.9} 0\\ \hline 
$l_{13}$&35&find&1&1&\cellcolor[gray]{0.9} 0\\ \hline 
$l_{14}$&47&grep&4&4&\cellcolor[gray]{0.9} 0\\ \hline 
$l_{15}$&5&tail&4&7&\cellcolor[gray]{0.9} -3\\ \hline 
$l_{16}$&12&cut&6&9&\cellcolor[gray]{0.9} -3\\ \hline 
$l_{17}$&46&grep&7&10&\cellcolor[gray]{0.9} -3\\ \hline 
$l_{18}$&7&seq&5&9&\cellcolor[gray]{0.9} -4\\ \hline 
$l_{19}$&15&du&1&6&\cellcolor[gray]{0.9} -5\\ \hline \hline
$l_{20}$&36&find&9&19&\cellcolor[gray]{0.9} -10\\ \hline 
$l_{21}$&37&find&22&34&\cellcolor[gray]{0.9} -12\\ \hline 
$l_{22}$&33&find&10&30&\cellcolor[gray]{0.9} -20\\ \hline 
$l_{23}$&2&od&86&186&\cellcolor[gray]{0.9} -100\\ \hline 
$l_{24}$&8&seq&22&160&\cellcolor[gray]{0.9} -138\\ \hline 
$l_{25}$&9&seq&10&157&\cellcolor[gray]{0.9} -147\\ \hline 
$l_{26}$&17&cut&37&195&\cellcolor[gray]{0.9} -158\\ \hline 
$l_{27}$&21&cut&20&191&\cellcolor[gray]{0.9} -171\\ \hline 
$l_{28}$&6&cut&5&206&\cellcolor[gray]{0.9} -201\\ \hline 
$l_{29}$&4&tail&28&431&\cellcolor[gray]{0.9} -403\\ \hline 
$l_{30}$&48&grep&26&497&\cellcolor[gray]{0.9} -471\\ \hline 
\end{tabular}
\end{table}

\section{Qualitative Analysis: Relevance of the Ranked SLOCs}
\label{sec:Qualitative}
To further understand our results, we manually investigate the produced rankings with respect to the studied faults.

To answer RQ3, we analysed the relevance of the faulty SLOC with the top ranked statements. We investigate the first $10$ SLOCs from each ranking according to the following procedure: (1) Manually locate each SLOC and the faulty SLOCs in the source code. (2) Guided by the faulty statements and dependancy analysis we assigned a relevance score ($)\sim 100$) to the SLOC. Such a relevance score depend on how fast a developer analysing the SLOC may end up to the faulty SLOC. We strictly considered dependencies and tried to quantify how easily a developer that starts from the provided statements and navigates through the dependencies can find the fault.  

The results of the relevance analysis are recorded in Table \ref{table:Relevance}. 
Note: Only faulty lines have a score of $100$; a score of $0$ do not necessarily mean that the SLOC is unrelated with the fault, but instead, means that it wouldn't quickly guide a developer to the faulty SLOCs according to our judgement. In situations where the fault is hard to localize and all SLOCs are given the same rank by the technique, we randomly choose to analyse $10$ SLOCs.

We observe that at least one relevant SLOC can be ranked high by the techniques for $80\%$ of the faults. Furthermore, The average relevance score of the top 10 results can be as high as $\sim 70/100$ for several faults. This results show that mutation-based FL techniques have the potential to help developer localize faults. They also point out a future direction of research that is to produce groups of statements that are related, as pointed out by Parnin and Orsro \cite{ParninO11}.

Finally, it is noted that the results presented here might be subjective (to some extend). Though, our goal was to give some hints regarding their relevance (with the studied faults) and not fully compare them. Furthermore, since we strictly considered dependencies all the statements with relevance higher than 0 are indeed related to the faults. 

\begin{table*}[t]
\tiny
\vspace{1.0em}
\centering
\caption{Relevance of the top 10 ranked statements. M indicates the average relevance.}
\label{table:Relevance}
\begin{tabular}{|l||*{11}{c|}>{\columncolor[gray]{0.9}}c||*{11}{c|}>{\columncolor[gray]{0.9}}c|} \hline
   & &\multicolumn{10}{c|}{\textbf{Top 10 Lines Relevance Scores $0\sim100$}}& & &\multicolumn{10}{c|}{\textbf{Top 10 Lines Relevance Scores $0\sim100$}}&  \\ \cline{3-12} \cline{15-24}
 &\multirow{-2}{*}{\textbf{\tiny ID}} & \textbf{1}&\textbf{2}&\textbf{3}&\textbf{4}&\textbf{5}&\textbf{6}&\textbf{7}&\textbf{8}&\textbf{9}&\textbf{10}&\multirow{-2}{*}{\textbf{\tiny M}} & 
 	\multirow{-2}{*}{\textbf{\tiny ID}} & \textbf{1}&\textbf{2}&\textbf{3}&\textbf{4}&\textbf{5}&\textbf{6}&\textbf{7}&\textbf{8}&\textbf{9}&\textbf{10}&\multirow{-2}{*}{\textbf{\tiny M}}\\ \hline \hline
 	
\textbf{\tiny Me}&&100&0&99&0&0&0&0&0&0&0&20&&0&0&0&85&85&85&0&0&0&0&26\\ \cline{3-12} \cline{15-24}
\textbf{\tiny MU}&\multirow{-2}{*}{1}&100&95&99&99&0&0&0&0&0&55&45&\multirow{-2}{*}{17}&0&0&0&0&0&0&0&0&0&0&0\\ \hline
\textbf{\tiny Me}&&0&0&0&0&0&0&0&0&0&0&0&&95&90&99&60&0&0&0&0&0&70&42\\ \cline{3-12} \cline{15-24}
\textbf{\tiny MU}&\multirow{-2}{*}{2}&0&0&0&0&0&0&0&0&0&0&0&\multirow{-2}{*}{18}&90&99&100&95&30&90&70&70&99&50&80\\ \hline
\textbf{\tiny Me}&&0&0&0&0&0&95&97&0&99&100&40&&100&96&0&100&0&0&97&95&0&0&49\\ \cline{3-12} \cline{15-24}
\textbf{\tiny MU}&\multirow{-2}{*}{3}&0&0&0&0&5&99&0&5&99&5&22&\multirow{-2}{*}{19}&100&100&0&0&0&95&20&0&20&2&34\\ \hline
\textbf{\tiny Me}&&0&0&0&0&0&0&0&0&0&0&0&&0&0&50&0&99&0&0&5&0&100&26\\ \cline{3-12} \cline{15-24}
\textbf{\tiny MU}&\multirow{-2}{*}{4}&0&0&0&0&0&0&0&0&0&0&0&\multirow{-2}{*}{20}&0&0&100&100&100&100&100&99&100&99&80\\ \hline
\textbf{\tiny Me}&&0&0&0&100&100&80&57&0&0&0&34&&5&5&0&0&0&0&0&0&0&0&1\\ \cline{3-12} \cline{15-24}
\textbf{\tiny MU}&\multirow{-2}{*}{5}&0&0&0&0&99&98&100&0&0&0&30&\multirow{-2}{*}{21}&100&80&75&60&60&20&15&0&0&0&41\\ \hline
\textbf{\tiny Me}&&0&0&0&0&100&0&99&99&99&0&40&&75&76&60&60&60&100&0&0&0&0&44\\ \cline{3-12} \cline{15-24}
\textbf{\tiny MU}&\multirow{-2}{*}{6}&0&0&0&0&0&0&0&0&0&0&0&\multirow{-2}{*}{22}&0&100&0&5&0&0&0&0&0&0&11\\ \hline
\textbf{\tiny Me}&&0&0&100&0&0&100&0&0&0&0&20&&0&20&20&20&20&50&50&0&0&100&28\\ \cline{3-12} \cline{15-24}
\textbf{\tiny MU}&\multirow{-2}{*}{7}&0&0&0&0&0&100&0&0&99&0&20&\multirow{-2}{*}{27}&100&20&20&20&20&0&100&50&50&70&45\\ \hline
\textbf{\tiny Me}&&0&0&50&0&85&90&90&95&99&40&55&&90&50&10&0&0&0&65&65&50&0&33\\ \cline{3-12} \cline{15-24}
\textbf{\tiny MU}&\multirow{-2}{*}{8}&99&0&50&85&90&85&90&95&90&0&69&\multirow{-2}{*}{32}&0&65&0&0&50&0&50&50&0&65&28\\ \hline
\textbf{\tiny Me}&&0&0&0&0&99&0&100&99&99&99&50&&50&0&90&50&0&0&0&0&0&100&29\\ \cline{3-12} \cline{15-24}
\textbf{\tiny MU}&\multirow{-2}{*}{9}&0&95&0&99&0&95&95&99&0&0&49&\multirow{-2}{*}{33}&0&50&0&0&0&0&0&0&90&0&14\\ \hline
\textbf{\tiny Me}&&0&100&100&95&0&0&0&0&0&0&30&&100&100&0&0&10&0&0&0&95&0&31\\ \cline{3-12} \cline{15-24}
\textbf{\tiny MU}&\multirow{-2}{*}{11}&100&100&0&0&0&0&100&0&0&0&30&\multirow{-2}{*}{35}&100&100&0&15&0&35&0&90&95&97&54\\ \hline
\textbf{\tiny Me}&&0&0&100&99&99&0&0&0&0&0&30&&100&0&0&100&0&0&0&0&0&0&20\\ \cline{3-12} \cline{15-24}
\textbf{\tiny MU}&\multirow{-2}{*}{12}&0&0&96&0&0&0&99&0&100&99&40&\multirow{-2}{*}{36}&3&0&90&0&0&3&0&0&0&0&10\\ \hline
\textbf{\tiny Me}&&0&0&0&0&0&80&0&0&0&0&8&&0&0&0&0&0&0&100&0&0&0&10\\ \cline{3-12} \cline{15-24}
\textbf{\tiny MU}&\multirow{-2}{*}{13}&0&0&0&0&0&0&0&0&0&0&0&\multirow{-2}{*}{37}&0&0&0&0&0&0&0&0&0&10&1\\ \hline
\textbf{\tiny Me}&&0&80&99&90&99&100&0&0&0&95&57&&0&0&0&0&0&100&100&100&0&99&40\\ \cline{3-12} \cline{15-24}
\textbf{\tiny MU}&\multirow{-2}{*}{14}&0&80&90&90&100&0&0&0&0&0&36&\multirow{-2}{*}{46}&0&0&0&0&0&0&0&0&0&100&10\\ \hline
\textbf{\tiny Me}&&100&99&40&68&40&95&95&90&87&0&72&&0&0&0&0&0&0&0&0&15&0&2\\ \cline{3-12} \cline{15-24}
\textbf{\tiny MU}&\multirow{-2}{*}{15}&40&99&95&0&0&100&0&0&0&0&34&\multirow{-2}{*}{47}&0&0&0&0&0&0&0&0&0&0&0\\ \hline
\textbf{\tiny Me}&&0&0&0&0&0&0&0&0&0&0&0&&0&0&0&0&0&0&0&87&70&0&16\\ \cline{3-12} \cline{15-24}
\textbf{\tiny MU}&\multirow{-2}{*}{16}&0&0&0&0&0&0&0&0&0&0&0&\multirow{-2}{*}{48}&0&0&0&0&0&0&0&0&0&0&0\\ \hline

\end{tabular}
\end{table*}

\section{Threats to Validity}
\label{threats}
Threats related to construct validity concern whether the experimental setup and metrics reflect real-world situation. 
In this study, a first thread to construct validity is due to the metrics used to measure that faults are localized. Following the recommendations made by literature \cite{Le:MAP-TopN}, we employed three metrics, i.e., Top N, MAP and MPS. The second threat to validity regards the definition of \textit{Faulty Line}. To reduce this threat, following existing studies \cite{Le:MAP-TopN}, we used the executable lines of code changed after bug-fixing commit on the buggy programs. However, it is possible that a fault can be fixed  by making changes many different locations. In order to further reduce this threat, more analysis is required on the faulty programs to gather more information about the root cause of the fault and therefore accurately evaluate the ranking of each fault localization technique.
The major threats to internal validity are related to uncontrolled factors that may affect our results. In this study, the major threats to internal validity are the tools we used and our implementation of the FL-techniques, data collection and analysis scripts. For this matter we employ tool that have previously been used in the academic software testing community. We did our best to resolve all errors from the tools that we implemented.
The major concern of external validity are whether the results of the present study can generalize to any other setting. The present study considered real faults selected by CoREBench's authors \cite{BohmeR14} from widely used programs, and the test case written by the programs' developers. The diversity and complexity of the faults as stated by CoREBench's authors is a leading factor contributing to the generalization of the present study on other C language programs. Whether this study is applicable on program written in other languages is unclear.  

\section{Related Work}
\label{sec:Related-Work}

In literature there is extensive research on both fault localization and mutation analysis. Despite this, only recently researchers attempted to combine them. The present section provides a brief description of the most relevant fault localization \ref{FL}, mutation-based fault localization \ref{MBFL} and mutation analysis methods \ref{Mutation}. 

\subsection{Spectrum-based fault localization}
\label{FL}

Spectrum-based fault localization has been widely studied. Perhaps the most popular and known technique is the Tarantula \cite{Jones:Score}. It uses program statements and a similarity function that relates program statements with faulty executions. A widely known improvement of this technique was the use of  Ochiai formula, which was demonstrated to be significantly more accurate  \cite{Abreu2007}. Later fault localization methods used other spectra types like program branches and du-pairs \cite{SantelicesJYH09}. All these approaches were unified by Santelices et al. \cite{SantelicesJYH09} who also showed that there is no certain type of spectra that can be stated as the most effective. Other researchers attempted to use different suspicious formulas  in order to improve effectiveness. Wong et al. \cite{WongDGL14} experimented with several heuristics and demonstrated that they could be better than Tarantula. Xie et al. \cite{XieCKX13} developed a theoretical framework and showed that some formulas are at least as good as others. 

All the above techniques rely on coverage spectra which have been shown to be ineffective compared to mutation-based approaches. For instance, MUSE was shown to be 25 times more precise than statement-based fault localization. Papadakis and Le Traon \cite{PapadakisT14}  also showed that Metallaxis outperforms all the different spectra types, i.e., statements, branches and du-pairs.

\subsection{Mutation-based fault localization}
\label{MBFL}

Research on mutation-based fault localization started by the work of Papadakis and Le Traon \cite{PapadakisT12, Papadakis:Metallaxis-FL} on Metallaxis. This work was then extended to reduce its application cost, using mutant-reduction techniques such as mutant sampling and selective mutation \cite{PapadakisT14}. The method was evaluated using the Proteum/FL tool \cite{PapadakisDT13} on hand-seeded faults and it was shown to be significantly more effective than the main spectrum-based fault localization techniques \cite{PapadakisT14}. A follow up work and probably the most recent one that attempted to improve Metallaxis was MUSE \cite{Moon:MUSE}. However, as already explained MUSE was evaluated on hand-seeded faults and was not compared with Metallaxis. Additionally, since it observes whether a mutant is turns a test case from passing to failing and vice versa, it is limited to mutants fixing the bugs, while it completely ignores the killed mutants by failing test cases. These issues motivated the present paper, which compares the effectiveness of Metallaxis and MUSE on real-world settings.  

Other mutation-based fault localization methods are due to Zhang et al. \cite{Zhang0K13} who used mutants to locate faults on evolving programs. Thus, the work of Zhang et al. was the first that aimed at ranking suspicious program edits by combining the information provided by program changes and the behaviour of mutants. Another similar work of this types is due to Murtaza et al. \cite{MurtazaHMG14} who used the impact of mutants on test execution in an attempt to produce traces that are similar to the (unknown) faulty ones. These traces can then be used to highly functions that are related to a given field failure. These two methods, although related, have a different goal than in our paper (Zhang et al. aims at program evolution and Murtaza et al. aims at field failures). In the present paper we aim at comparing and assessing the MUSE and Metallaxis on real-world settings given only the program with its passing and failing test cases.

\subsection{Mutation Analysis}
\label{Mutation}

Mutation analysis was suggested as a way to solicit high quality test cases \cite{Offutt11}. Then, researchers extended to support many software engineering tasks such as suggesting test oracles \cite{FraserZ12}, reducing test suites \cite{GligoricNLM14}, and debugging activities like fault localization \cite{Papadakis:Metallaxis-FL} and bug fixing \cite{DebroyW14}. The idea  has also been applied on several types of models, e.g., feature models \cite{HenardPPKT13}, combinatorial interaction testing models \cite{PapadakisHT14} and behavioral models \cite{DevroeyPPLSH16, AichernigBJKST15}. 

Mutation is a popular technique mainly because it can support experimentation \cite{AndrewsBLN06} and can subsume most of the other software testing techniques \cite{0020331}. Recently, many robust and fast mutation testing tools have been built and integrated with software development tools. Thus, making it easy to use. One of the main problems of mutation is the equivalent mutants. Despite the recent success in automating this process \cite{Papadakis:Trivial}, the problem remains. Luckily, equivalent mutants are not killed and thus, they do not pose any problems on fault localization.

\section{Conclusion}
\label{sec:Conclusion}

This paper empirically investigates a relatively new direction of research, the mutation-based fault localization. 
The paper provides promising results indicating that by inspecting only 10 statements developers can successfully locate approximately 50\% of the studied faults. The number of located faults increases to 80\% for developers inspecting 25 statements. Overall,  Metallaxis and MUSE can successfully locate the studied faults by respectively investigating the 18\% and 37\% of the executable statements of the programs under analysis. Therefore, indicating that Metallaxis significantly outperforms MUSE.

Another important finding of our results is the relation of the top-ranked statements with the faults under analysis. Our results show that at least one relevant statement exist in the top 10 statements for 80\% of the studied faults. Furthermore, the average relevance of the top 10 statements of Metallaxis and MUSE is 28.16\% and 25.88\%, respectively. This is important since it ensures that developers will maintain their interest when using fault localization \cite{ParninO11}.

Overall, our results indicate that mutation-based fault localization achieves to successfully highlight faulty statements. It also points out several closely related, to the sought faults, statements providing evidence that the produced rankings can be of practical value. Also, our results indicate that Metallaxis produces twice more accurate results than MUSE.



%
\balance
\bibliographystyle{abbrv}
\bibliography{references.bib}  

\begin{thebibliography}{10}

\bibitem{Abreu2007}
R.~Abreu, P.~Zoeteweij, and A.~J.~C. van Gemund.
\newblock On the accuracy of spectrum-based fault localization.
\newblock In {\em Proceedings of the Testing: Academic and Industrial
  Conference Practice and Research Techniques - MUTATION}, TAICPART-MUTATION
  '07, pages 89--98, 2007.

\bibitem{AichernigBJKST15}
B.~K. Aichernig, H.~Brandl, E.~J{\"{o}}bstl, W.~Krenn, R.~Schlick, and
  S.~Tiran.
\newblock Killing strategies for model-based mutation testing.
\newblock {\em Softw. Test., Verif. Reliab.}, 25(8):716--748, 2015.

\bibitem{0020331}
P.~Ammann and J.~Offutt.
\newblock {\em Introduction to software testing}.
\newblock Cambridge University Press, 2008.

\bibitem{AndrewsBLN06}
J.~H. Andrews, L.~C. Briand, Y.~Labiche, and A.~S. Namin.
\newblock Using mutation analysis for assessing and comparing testing coverage
  criteria.
\newblock {\em {IEEE} Trans. Software Eng.}, 32(8):608--624, 2006.

\bibitem{BakerH13}
R.~Baker and I.~Habli.
\newblock An empirical evaluation of mutation testing for improving the test
  quality of safety-critical software.
\newblock {\em IEEE Transactions on Software Engineering}, 39(6):787--805,
  2013.

\bibitem{BohmeR14}
M.~B{\"{o}}hme and A.~Roychoudhury.
\newblock Corebench: studying complexity of regression errors.
\newblock In {\em International Symposium on Software Testing and Analysis,
  {ISSTA} '14, San Jose, CA, {USA} - July 21 - 26, 2014}, pages 105--115, 2014.

\bibitem{DebroyW14}
V.~Debroy and W.~E. Wong.
\newblock Combining mutation and fault localization for automated program
  debugging.
\newblock {\em Journal of Systems and Software}, 90:45--60, 2014.

\bibitem{DevroeyPPLSH16}
X.~Devroey, G.~Perrouin, M.~Papadakis, A.~Legay, P.~Schobbens, and P.~Heymans.
\newblock Featured model-based mutation analysis.
\newblock In {\em Proceedings of the 38th International Conference on Software
  Engineering, {ICSE} 2016, Austin, TX, USA, May 14-22, 2016}, pages 655--666,
  2016.

\bibitem{FraserZ12}
G.~Fraser and A.~Zeller.
\newblock Mutation-driven generation of unit tests and oracles.
\newblock {\em {IEEE} Trans. Software Eng.}, 38(2):278--292, 2012.

\bibitem{GligoricNLM14}
M.~Gligoric, S.~Negara, O.~Legunsen, and D.~Marinov.
\newblock An empirical evaluation and comparison of manual and automated test
  selection.
\newblock In {\em {ACM/IEEE} International Conference on Automated Software
  Engineering, {ASE} '14, Vasteras, Sweden - September 15 - 19, 2014}, pages
  361--372, 2014.

\bibitem{CPHJT16}
C.~Henard, M.~Papadakis, M.~Harman, Y.~Jia, and Y.~L. Traon.
\newblock Comparing white-box and black-box test prioritization.
\newblock In {\em {ICSE}}, 2016.

\bibitem{HenardPPKT13}
C.~Henard, M.~Papadakis, G.~Perrouin, J.~Klein, and Y.~L. Traon.
\newblock Assessing software product line testing via model-based mutation: An
  application to similarity testing.
\newblock In {\em Sixth {IEEE} International Conference on Software Testing,
  Verification and Validation, {ICST} 2013 Workshops Proceedings, Luxembourg,
  Luxembourg, March 18-22, 2013}, pages 188--197, 2013.

\bibitem{Jeffrey:2008:FLU}
D.~Jeffrey, N.~Gupta, and R.~Gupta.
\newblock Fault localization using value replacement.
\newblock In {\em Proceedings of the 2008 International Symposium on Software
  Testing and Analysis}, ISSTA '08, pages 167--178, New York, NY, USA, 2008.
  ACM.

\bibitem{Jones:Score}
J.~A. Jones and M.~J. Harrold.
\newblock Empirical evaluation of the tarantula automatic fault-localization
  technique.
\newblock In {\em Proceedings of the 20th IEEE/ACM International Conference on
  Automated Software Engineering}, ASE '05, pages 273--282, New York, NY, USA,
  2005. ACM.

\bibitem{Le:MAP-TopN}
T.-D.~B. Le, R.~J. Oentaryo, and D.~Lo.
\newblock Information retrieval and spectrum based bug localization: Better
  together.
\newblock In {\em Proceedings of the 2015 10th Joint Meeting on Foundations of
  Software Engineering}, ESEC/FSE 2015, pages 579--590, New York, NY, USA,
  2015. ACM.

\bibitem{Manning:MAP}
C.~D. Manning, P.~Raghavan, and H.~Sch\"{u}tze.
\newblock {\em Introduction to Information Retrieval}.
\newblock Cambridge University Press, New York, NY, USA, 2008.

\bibitem{MasriA14}
W.~Masri and R.~A. Assi.
\newblock Prevalence of coincidental correctness and mitigation of its impact
  on fault localization.
\newblock {\em {ACM} Trans. Softw. Eng. Methodol.}, 23(1):8:1--8:28, 2014.

\bibitem{Moon:MUSE}
S.~Moon, Y.~Kim, M.~Kim, and S.~Yoo.
\newblock Ask the mutants: Mutating faulty programs for fault localization.
\newblock In {\em Proceedings of the 2014 IEEE International Conference on
  Software Testing, Verification, and Validation}, ICST '14, pages 153--162,
  Washington, DC, USA, 2014. IEEE Computer Society.

\bibitem{MurtazaHMG14}
S.~S. Murtaza, A.~Hamou{-}Lhadj, N.~H. Madhavji, and M.~Gittens.
\newblock An empirical study on the use of mutant traces for diagnosis of
  faults in deployed systems.
\newblock {\em Journal of Systems and Software}, 90:29--44, 2014.

\bibitem{OffuttH96}
A.~J. Offutt and J.~H. Hayes.
\newblock A semantic model of program faults.
\newblock In {\em {ISSTA}}, pages 195--200, 1996.

\bibitem{Offutt11}
J.~Offutt.
\newblock A mutation carol: Past, present and future.
\newblock {\em Information {\&} Software Technology}, 53(10):1098--1107, 2011.

\bibitem{PadioleauLHM08}
Y.~Padioleau, J.~L. Lawall, R.~R. Hansen, and G.~Muller.
\newblock Documenting and automating collateral evolutions in linux device
  drivers.
\newblock In {\em Proceedings of the 2008 EuroSys Conference, Glasgow,
  Scotland, UK, April 1-4, 2008}, pages 247--260, 2008.

\bibitem{PapadakisDT13}
M.~Papadakis, M.~E. Delamaro, and Y.~L. Traon.
\newblock Proteum/fl: {A} tool for localizing faults using mutation analysis.
\newblock In {\em 13th {IEEE} International Working Conference on Source Code
  Analysis and Manipulation, {SCAM} 2013, Eindhoven, Netherlands, September
  22-23, 2013}, pages 94--99, 2013.

\bibitem{PapadakisHT14}
M.~Papadakis, C.~Henard, and Y.~L. Traon.
\newblock Sampling program inputs with mutation analysis: Going beyond
  combinatorial interaction testing.
\newblock In {\em Seventh {IEEE} International Conference on Software Testing,
  Verification and Validation, {ICST} 2014, March 31 2014-April 4, 2014,
  Cleveland, Ohio, {USA}}, pages 1--10, 2014.

\bibitem{Papadakis:Trivial}
M.~Papadakis, Y.~Jia, M.~Harman, and Y.~Le~Traon.
\newblock Trivial compiler equivalence: A large scale empirical study of a
  simple, fast and effective equivalent mutant detection technique.
\newblock In {\em Proceedings of the 37th International Conference on Software
  Engineering - Volume 1}, ICSE '15, pages 936--946, Piscataway, NJ, USA, 2015.
  IEEE Press.

\bibitem{Papadakis:Metallaxis-FL}
M.~Papadakis and Y.~Le~Traon.
\newblock Metallaxis-fl: Mutation-based fault localization.
\newblock {\em Softw. Test. Verif. Reliab.}, 25(5-7):605--628, Aug. 2015.

\bibitem{PapadakisT12}
M.~Papadakis and Y.~L. Traon.
\newblock Using mutants to locate "unknown" faults.
\newblock In {\em Fifth {IEEE} International Conference on Software Testing,
  Verification and Validation, {ICST} 2012, Montreal, QC, Canada, April 17-21,
  2012}, pages 691--700, 2012.

\bibitem{PapadakisT14}
M.~Papadakis and Y.~L. Traon.
\newblock Effective fault localization via mutation analysis: a selective
  mutation approach.
\newblock In {\em Symposium on Applied Computing, {SAC} 2014, Gyeongju,
  Republic of Korea - March 24 - 28, 2014}, pages 1293--1300, 2014.

\bibitem{ParninO11}
C.~Parnin and A.~Orso.
\newblock Are automated debugging techniques actually helping programmers?
\newblock In {\em Proceedings of the 20th International Symposium on Software
  Testing and Analysis, {ISSTA} 2011, Toronto, ON, Canada, July 17-21, 2011},
  pages 199--209, 2011.

\bibitem{Saha:TopN}
R.~K. Saha, M.~Lease, S.~Khurshid, and D.~E. Perry.
\newblock Improving bug localization using structured information retrieval.
\newblock In {\em Automated Software Engineering (ASE), 2013 IEEE/ACM 28th
  International Conference on}, pages 345--355, Nov 2013.

\bibitem{SantelicesJYH09}
R.~A. Santelices, J.~A. Jones, Y.~Yu, and M.~J. Harrold.
\newblock Lightweight fault-localization using multiple coverage types.
\newblock In {\em 31st International Conference on Software Engineering, {ICSE}
  2009, May 16-24, 2009, Vancouver, Canada, Proceedings}, pages 56--66, 2009.

\bibitem{WongDGL14}
W.~E. Wong, V.~Debroy, R.~Gao, and Y.~Li.
\newblock The dstar method for effective software fault localization.
\newblock {\em {IEEE} Trans. Reliability}, 63(1):290--308, 2014.

\bibitem{XieCKX13}
X.~Xie, T.~Y. Chen, F.~Kuo, and B.~Xu.
\newblock A theoretical analysis of the risk evaluation formulas for
  spectrum-based fault localization.
\newblock {\em {ACM} Trans. Softw. Eng. Methodol.}, 22(4):31, 2013.

\bibitem{XuanM14}
J.~Xuan and M.~Monperrus.
\newblock Test case purification for improving fault localization.
\newblock In {\em Proceedings of the 22nd {ACM} {SIGSOFT} International
  Symposium on Foundations of Software Engineering, (FSE-22), Hong Kong, China,
  November 16 - 22, 2014}, pages 52--63, 2014.

\bibitem{Zhang0K13}
L.~Zhang, L.~Zhang, and S.~Khurshid.
\newblock Injecting mechanical faults to localize developer faults for evolving
  software.
\newblock In {\em Proceedings of the 2013 {ACM} {SIGPLAN} International
  Conference on Object Oriented Programming Systems Languages {\&}
  Applications, {OOPSLA} 2013, part of {SPLASH} 2013, Indianapolis, IN, USA,
  October 26-31, 2013}, pages 765--784, 2013.

\bibitem{Zhou:TopN}
J.~Zhou, H.~Zhang, and D.~Lo.
\newblock Where should the bugs be fixed? - more accurate information
  retrieval-based bug localization based on bug reports.
\newblock In {\em Proceedings of the 34th International Conference on Software
  Engineering}, ICSE '12, pages 14--24, Piscataway, NJ, USA, 2012. IEEE Press.

\end{thebibliography}
\end{document}